\documentclass[11pt,twoside]{article}
\usepackage{graphicx,epsfig,natbib,epstopdf}
\usepackage{CS18}
\begin{document}
%
%
%
\title{MHD simulations of near-surface convection in cool
main-sequence stars}
%
%
\author{Benjamin Beeck$^{1}$, Manfred Sch{\"u}ssler$^{1}$, Ansgar Reiners$^{2}$}
\affil{$^1$Max Planck Institute for Solar System Research,
  Justus-von-Liebig-Weg 3, 37077 G{\"o}ttingen, Germany}
\affil{$^2$Institute for Astrophysics, University of G{\"o}ttingen,
  Friedrich-Hund-Platz 1, 37077 G{\"o}ttingen, Germany}
\begin{abstract}
%
%
The solar photospheric magnetic field is highly structured owing to its
interaction with the convective flows. Its local structure has a strong
influence on the profiles of spectral lines not only by virtue of the Zeeman
effect, but also through the modification of the thermodynamical structure
(e.g. line weakening in hot small-scale magnetic structures). Many stars
harbor surface magnetic fields comparable to or larger than the Sun at solar
maximum. Therefore, a strong influence of the field on the surface convection
and on spectral line profiles can be expected.\par

We carried out 3D local-box MHD simulations of unipolar magnetized regions
(average fields of 20, 100, and 500G) with parameters corresponding to six
main-sequence stars (spectral types F3V to M2V). The influence of the magnetic
field on the convection and the local thermodynamical structure were analyzed
in detail. For three spectral lines, we determined the impact of the magnetic
field on the disc-integrated Stokes-$I$ profiles. Line weakening has in many
cases a stronger impact on the spectral line profiles than the Zeeman
effect. Moreover, for some stars, the correlation between the magnetic field
and the vertical velocity strongly influences the line shapes. These effects
can impair determinations of stellar magnetic fields since currently used methods
neglect the local structure of the magnetic field\index{local magnetic field structure} and its interaction with the
convective flows. The MHD simulations presented can be used to quantify these
effects and thus help to improve magnetic field measurements of cool
main-sequence stars.

\end{abstract}
%
%
%
%
%
\section{Introduction}

The interaction between magnetic fields and convective flows plays an
important role in the convectively unstable outer layers of cool main-sequence stars. In the Sun, many local effects of the magnetic field (sunspots, faculae, chromospheric
activity, etc.) can be observed in great detail. For other stars such spatially highly
resolved observations are, however, not possible.\par

In recent years, many successful magnetic field detections in cool
main-sequence stars exploiting the Zeeman effect have been reported
\citep[see][for a review]{Reiners12}. With Zeeman Doppler imaging (ZDI) it has
become possible to extract the large-scale geometry of the magnetic field of
some rapidly rotating cool stars \citep[see, e.\,g.,][]{ZDI1}. However, for most measurements
of stellar magnetic fields the spatial correlations between the velocity
field, temperature, pressure, and density with the magnetic field have to be
ignored as they are unknown. In the Sun, the magnetic field in the photosphere
is concentrated in regions, which can be much brighter or much darker
(i.\,e., cooler and denser) than the quiet Sun. It is thus likely that the magnetic field has a similar local
impact on the photospheres of other stars. As the spectral line profiles which
are used to determine the magnetic field are also sensitive to thermodynamic
quantities (temperature, pressure, etc.), their disk-integrated profiles
including the Zeeman broadening/splitting and the polarization are affected by the correlation between thermodynamic properties and the magnetic
field. For example, \citet{Rosen12} showed that the ZDI reconstruction of the
magnetic field from artificial spectroscopic data fails if the magnetic field
is concentrated in dark spots.\par 

We have run three-dimensional magnetohydrodynamic\index{magnetohydrodynamics} simulations\index{simulations} in order to
investigate the details of the magnetoconvection\index{magnetoconvection} in the near surface layers of
cool main-sequence stars\index{main-sequence stars}. We analyze the impact of a moderate unipolar
magnetic field\index{magnetic field} (20 to 500\,G horizontally averaged flux density) on the convection\index{convection} and on the disk-integrated
profiles of three sample spectral lines.

\section{MHD Simulations with MURaM}

\subsection{MURaM code}
The MURaM code\index{MURaM code} solves the fully compressible MHD
equations on a three-dimensional Cartesian grid with explicit time
stepping. Radiative transport is approximated with the opacity binning method
 ($\tau$ sorting for the individual atmospheres) with four bins resolved in three
ray directions per octant\citep[][and references therein]{MURaM3}. The code
uses the OPAL equation of state
\citep{OPAL} which includes partial ionization. Opacities and equation of
state used are for the solar chemical composition published by \citet{AndersGrevesse89}. For more details on the code,
 see \citet{MURaM1} and \citet{MURaM2}; for a comparison between MURaM and two
 similar codes for solar parameters see \citet{Codecomparison}. The version of
 the code used for the simulations considered here, uses the slope
 limiter and the modified Lorentz force introduced by \citet{Rempel09}. 

\subsection{Simulation setup}
We simulated the near-surface convection with the properties resembling those
of six cool main-sequence stars that roughly correspond to spectral types F3V,
G2V (Sun), K0V, K5V, M0V, and M2V. The sizes of the computational domain were
adapted to the horizontal scale of the convection (granulation) and to the
pressure scale heights \citep[see][]{Paper1}. As initial condition and as a
reference non-magnetic (hydrodynamic) simulations were run. These were
analyzed in detail in \citet{Paper1} and \citet{Paper2}. For each spectral
type three magnetic simulations were run. The initial condition for the
magnetic field in these three runs was a vertical homogeneous field of 20,
100, and 500\,G, respectively. After an initial restructuring phase lasting about 10--30
stellar minutes, the simulations reached statistically stationary states. The snapshots
considered were all taken from this ``relaxed'' stationary phase,
about 1--2 h (stellar time) after the magnetic field was introduced \citep[for more details, see][]{diss}.

\section{Results}

\subsection{Surface magnetoconvection}
\begin{figure}
\includegraphics[width=0.31\textwidth]{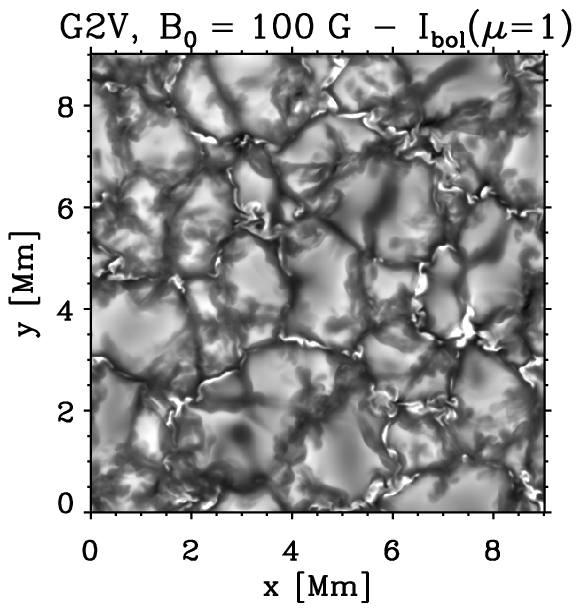}~\includegraphics[width=0.31\textwidth]{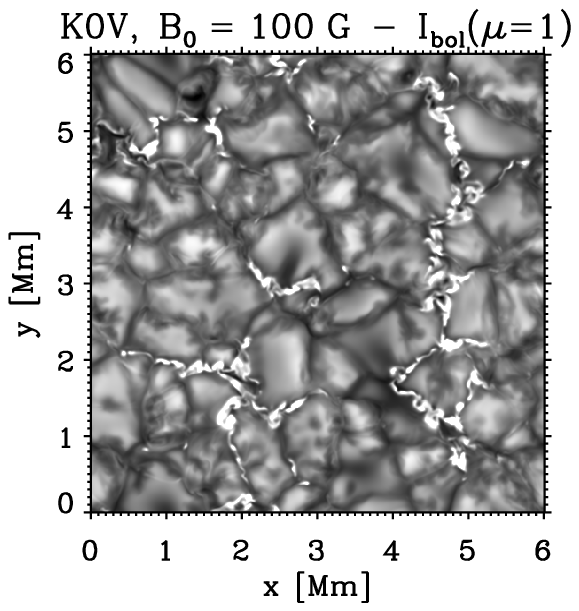}~\includegraphics[width=0.31\textwidth]{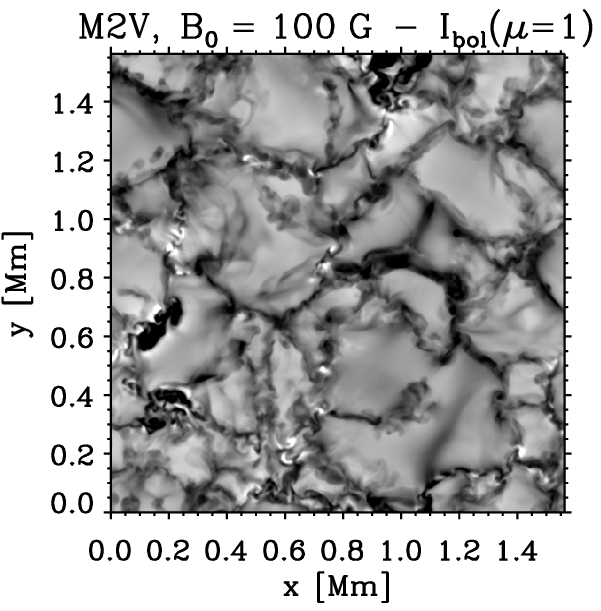}\\
\includegraphics[width=0.31\textwidth]{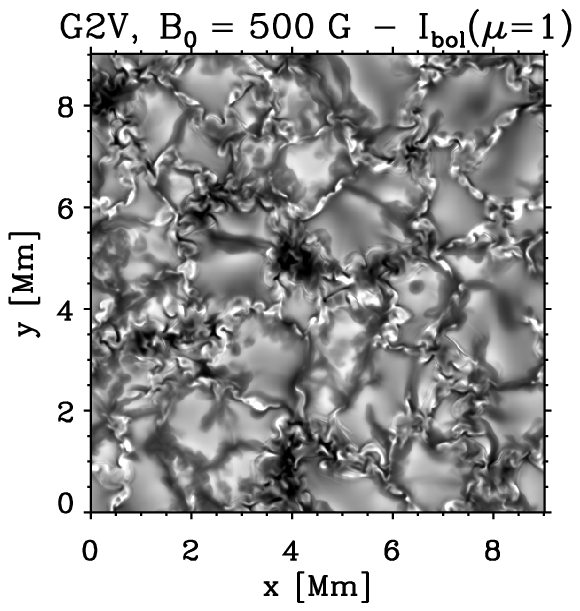}~\includegraphics[width=0.31\textwidth]{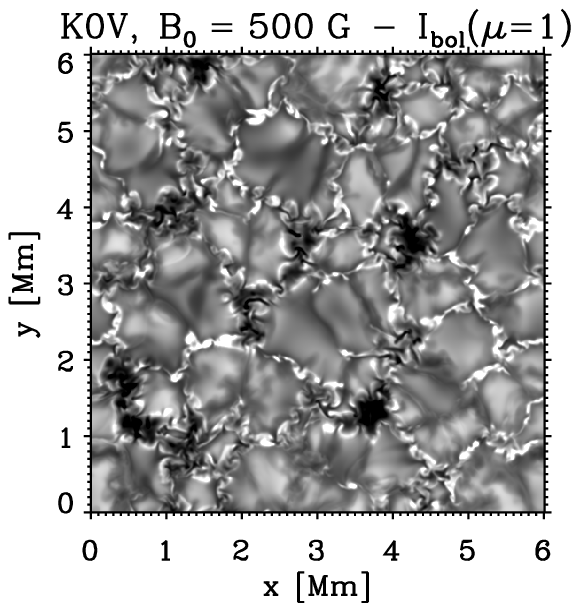}~\includegraphics[width=0.31\textwidth]{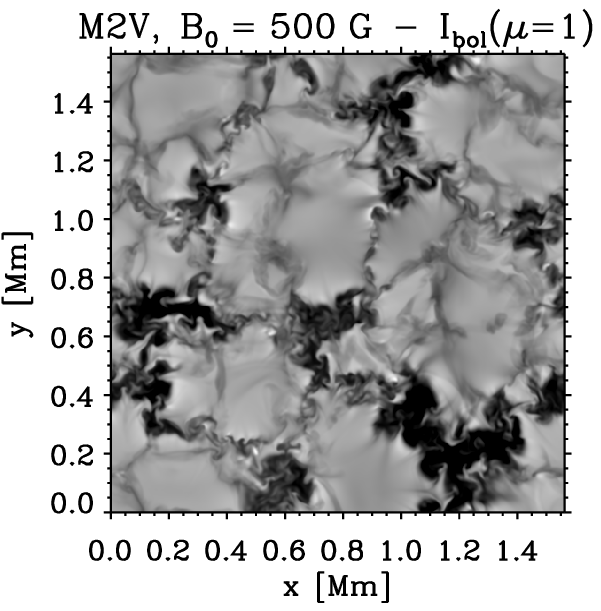}\\
\caption{Bolometric intensity maps (at vertical view,
  i.\,e. $\mu=\cos\theta=1$) for one snapshot of six of the MHD
  simulations. The approximate spectral types and the mean vertical magnetic
  field are given in the titles of each panel. The gray scale saturates at
  $\pm 2.5$ standard deviations from the mean intensity.}\label{vert_int}
\end{figure}
Figure~\ref{vert_int} shows the vertical bolometric intensity of six of the
magnetic runs (G2V, K0V, and M2V, each with 100 and 500\,G average vertical
magnetic field strength). The magnetic field
is advected into the intergranular lanes where it accumulates and forms
structures with a local field strength of roughly 1.5--2.5\,kG, which are visible in the intensity images
because they have either reduced or enhanced intensity. In general, smaller
magnetic flux concentrations are bright and
larger ones are dark. As the
evacuation caused by the magnetic pressure in these structures produces
depressions of the optical surface (or enhances the already existing depressions in the downflow lanes), hot side walls form around the magnetic flux concentrations. These can efficiently
heat small structures (large side-wall area compared to the volume of the
structure). If these depressions become much wider than they are deep, the
sidewall heating is less efficient. The structures then become dark in the
absence of sufficient convective heating, because magnetic flux concentrations
suppress horizontal inflows owing to the strong vertical magnetic field. Figure~\ref{vert_int} also illustrates that there is a
substantial difference between M dwarfs and the hotter main-sequence stars. While in the
hotter stars the bright magnetic features are very pronounced even at an average field of
500\,G, in the M stars there are only very few bright structures and many more
dark ones. The depressions in the optical surface of magnetic flux concentraions in these stars are much
shallower compared to their horizontal size than on hotter stars. \citet{diss} shows that these structures are essentially at
rest and have an intensity of only 80--90\% of the mean intensity while the
typical spatial variation of the intensity is only 2--3\% in the non-magnetic M star simulations. 

\subsection{Zeeman effect and thermodynamic effects of the magnetic field}

\begin{figure}
\includegraphics[width=0.57\textwidth]{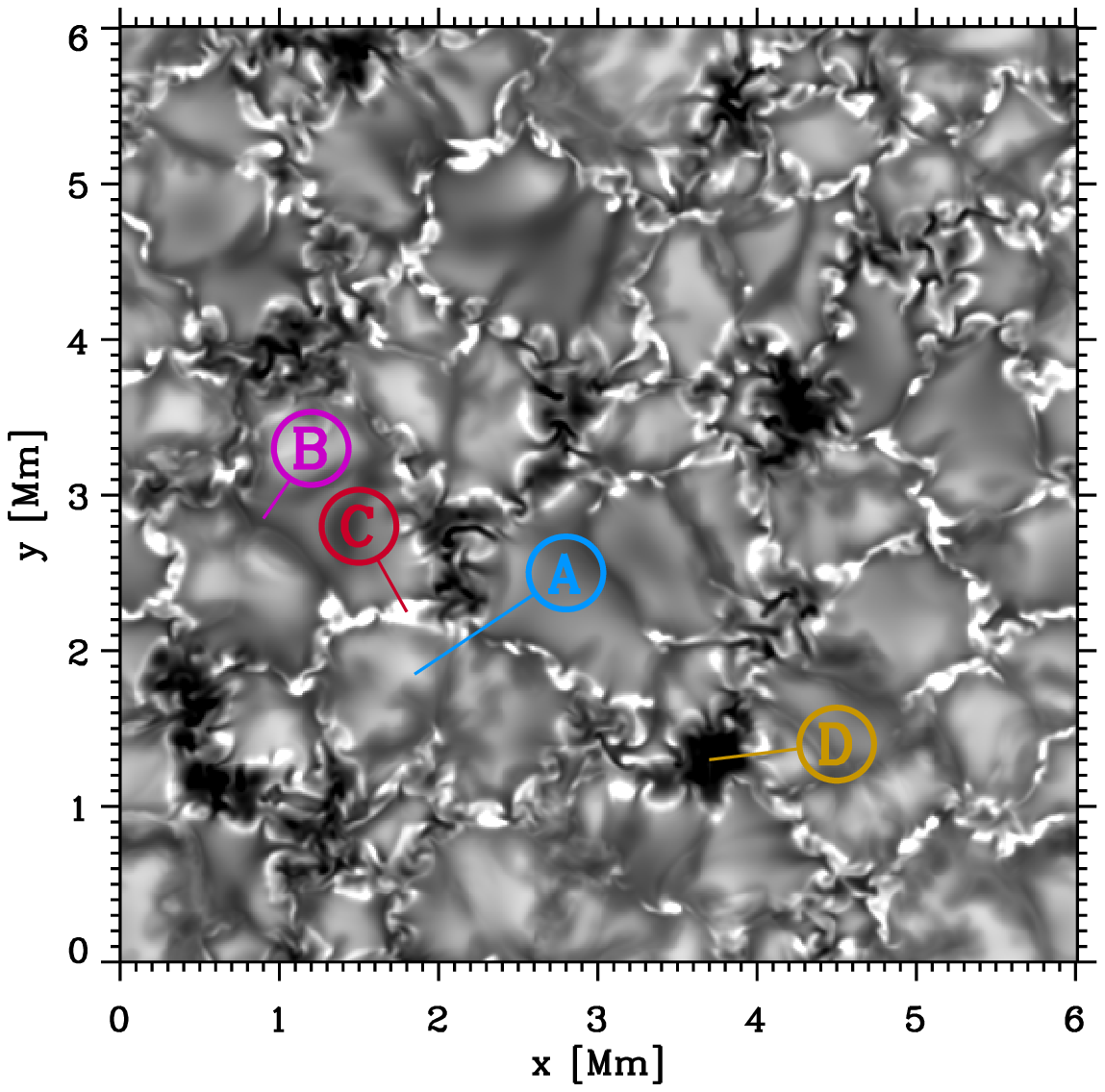}~~~~\parbox[b]{0.38\paperwidth}{\includegraphics[width=0.37\textwidth]{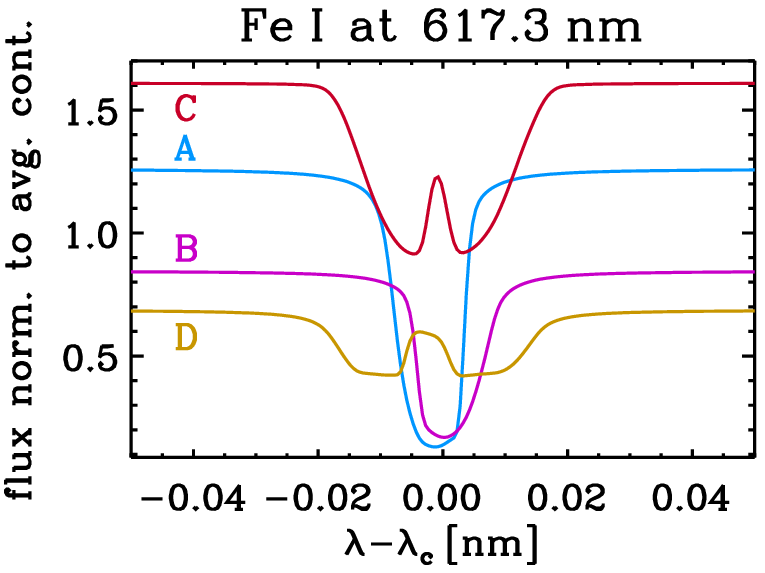}\\\includegraphics[width=0.37\textwidth]{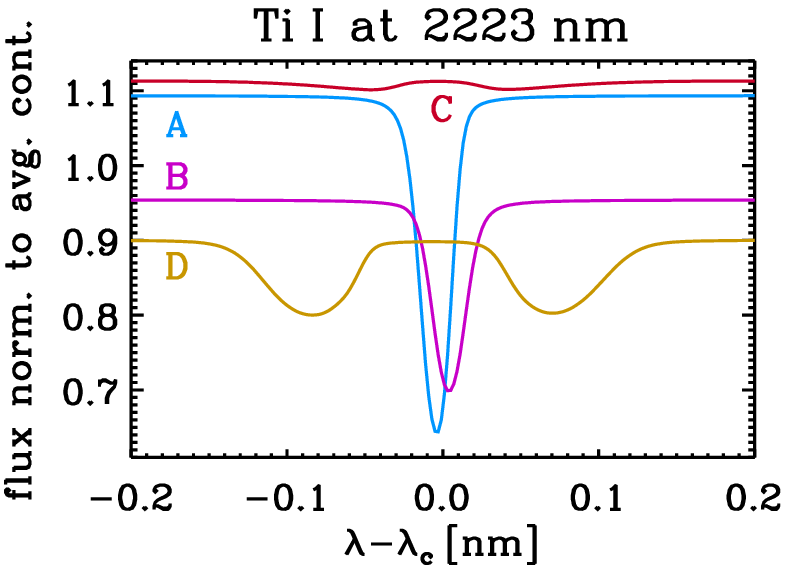}}
\caption{Local spectral line profiles. Left panel: Bolometric intensity map of
  a snapshot from the K0V-star simulation with four points (A to D)
  marked. Right panels: Local spectral line profiles (normalized to the mean
  continuum level) at the four points marked in the left panel for two
  spectral line (top: {\rm Fe}\,\textsc{i} at 617.3\,nm; bottom: {\rm Ti}\,\textsc{i} at
  2223\,nm)}\label{local_spec}
\end{figure}

Figure~\ref{local_spec} compares the local profiles of two spectral lines
(Fe\,\textsc{i} at 617.3\,nm and Ti\,\textsc{i} at 2223\,nm) at four different
points ({\it A} to {\it D}) for a snapshot of the K0V star simulation with
500\,G average field. Two points ({\it A} and {\it B}) are weakly magnetized
points: {\it A} is part of an upflow, {\it B} is situated in a non-magnetic
downflow. Consequently, the spectral line is shifted to the blue in {\it A} and
to the red in {\it B}. The different temperatures at these two points result
in different line depths and different continuum levels. The depth dependence of the line-of-sight velocity affects the line
profiles in the form of asymmetries\index{spectral line asymmetries} (especially for the strongly saturated Fe\,\textsc{i}
line). The other two points are located in magnetic structures: {\it C} is the
center of a small, bright structure, {\it D} is in the darkest part of an
extended dark structure. As both lines have high Land{\'e} factors (2.5 and
1.6, respectively), the Zeeman
effect\index{Zeeman effect} splits the line profiles into its Zeeman components. The
height-dependent magnetic field is responsible for asymmetries of the
individual components. The equivalent width of the Fe\,\textsc{i} line is similar in {\it C} and {\it D}. In a disk-integrated line profile,
{\it D} will, however, have a lower weight than {\it C} as the integrated (absorbed) flux
of the spectral line is lower due to the strong difference in continuum
level. The Ti\,\textsc{i} line is very sensitive to the temperature in the
typical temperature regime of the K0V star photosphere (mostly owing to
ionization of titanium). This leads to a substantial
weakening of the Ti\,\textsc{i} line in bright (=hot) structures: the
equivalent width of the Ti\,\textsc{i} line at {\it C} is strongly reduced to
only a few percent of its value at the other points. For this spectral line,
bright magnetic structures such as at {\it C} do not contribute much to the
disk-integrated line profiles.\par
\begin{figure}
\includegraphics[width=0.48\textwidth]{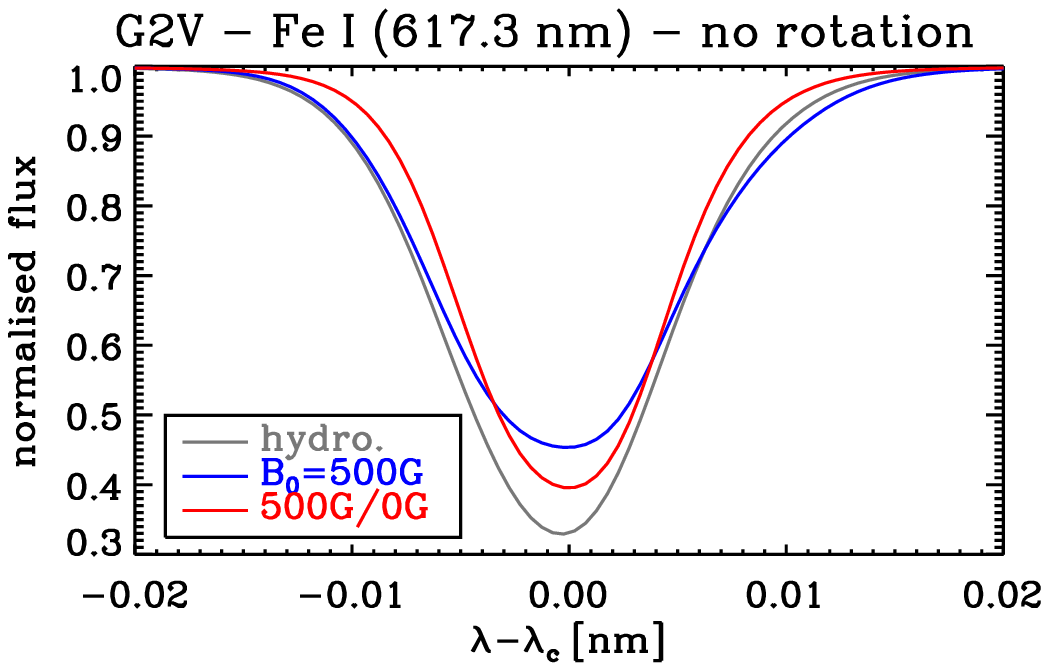}~~\includegraphics[width=0.48\textwidth]{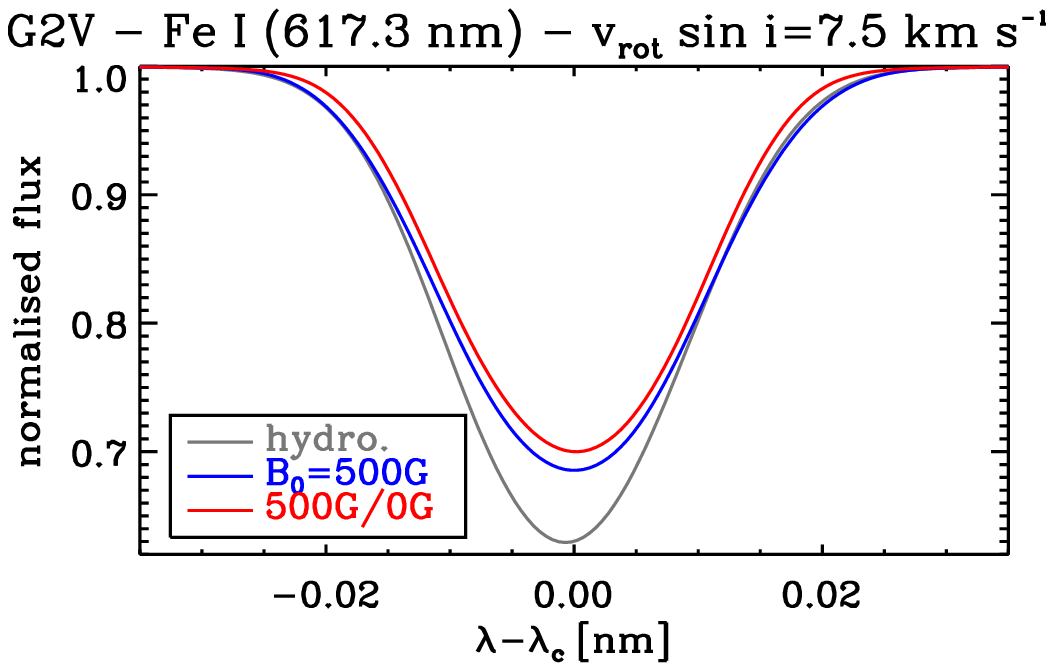}
\caption{Illustration of the relative effects of Zeeman effect and
  thermodynamic differences between non-magnetic and magnetic simulations
  (here: G2V) on the {\rm Fe}\,\textsc{i} line profile. The gray curve corresponds to
the average disk-integrated line profile of the non-magnetic simulation, the
blue curve corresponds to the average disk-integrated line profile of the
run with 500\,G average field. The red curve shows the line profile calculated
from the same atmosphere as the blue curve, but with the Zeeman effect ignored.}\label{Zeeman_vs_TD}
\end{figure}
As Figure~\ref{local_spec} shows, the formation of bright and dark magnetic
structures results in very different modifications of the local profiles of
the same spectral line in the same atmosphere. Figure~\ref{Zeeman_vs_TD} shows the disk-integrated profile of the
Fe\,\textsc{i} line at 617.3\,nm, without rotation (left panel) and with a solar-like
differential rotation with $v_{\mathrm{eq}}\sin i=7.5\,\mathrm{km\,s^{-1}}$ (right panel). The gray curves are
the average line profiles resulting from six snapshots of the non-magnetic
(hydrodynamic) run of the G2V star, while the blue curves are the
corresponding profiles resulting from six snapshots of the magnetohydrodynamic run
of the G2V star with 500\,G average field. The red curve results from the same
six snapshots (G2V, 500\,G), but for the line calculation the field was ignored
(i.\,e. the Zeeman effect was artificially turned off). The difference between
the blue and red curves is thus only due to the Zeeman effect, while the
thermodynamic local modifications of the atmosphere owing to the magnetic field
cause the difference between the gray and the red curve. Without rotation
(left panel), the thermodynamic modifications reduce the equivalent width and
introduce a relative redshift of the line. The reduced equivalent width results from the existence of bright magnetic structures. At solar
photospheric conditions, the Fe\,\textsc{i} line at 617.3\,nm is strongly temperature
sensitive and a similar effect of line weakening as for Ti\,\textsc{i} in the K0V star
(see point {\it C} in Fig.~\ref{local_spec}) can be observed. The relative redshift
can be explained by an increase of the filling factor of downflows and a
higher downflow speed at the same optical depth in the magnetic simulation
runs \citep[see][for more details]{diss}. The Zeeman effect broadens the line significantly and increases the
equivalent width (owing to reduced saturation). With a
rotation rate of 7.5\,km\,s$^{-1}$, which is a typical rotation rate for
stars somewhat more active than the Sun, the rotational broadening is,
however, already much stronger than the Zeeman broadening and the
thermodynamic effects clearly have a bigger impact on the line profile than
the Zeeman effect. Magnetic field measurements
neglecting the effects of the magnetic field on the convection probably suffer
from the impact of the difference in thermodynamic structure between the magnetized and the non-magnetized atmosphere.

\section{Conclusion}
\subsection{Discussion}

As illustrated by our 3D simulations, stellar magnetic fields interact with
convective flows and locally modify the convection as well as the
radiative properties of the stellar photospheres. As a consequence,
local concentrations of magnetic field have thermodynamic
properties which are very different from the surrounding atmosphere. The resulting
spectral line profiles do not only show a Zeeman splitting/broadening but also
the signatures of the modified thermodynamics. In some cases, line weakening\index{spectral line weakening}
(due to, e.\,g., ionization in the bright magnetic structures) can strongly
reduce the equivalent width of the lines in concentrations of magnetic
field. \citet{diss} shows that this results in a underestimation of $Bf$ (where $B$ is the field strength of the magnetic part of the star and $f$ is the area fraction covered by $B$) in a
two-component-model fit to the spectral line profiles if one neglects the
thermodynamic effects for the fit. In order to accurately measure the magnetic
field and other properties of a stellar atmosphere, the active role of the
magnetic field must be taken into account.

\subsection{Outlook}
So far, only unipolar regions of moderate average field strength of up to
500\,G have been simulated. The next step will be the simulation of regions
with stronger mean magnetic field (starspot umbrae), or more complex
geometries (e.\,g., bipolar regions). For the disk-integrated line profiles we
assumed the star to be homogeneous on large scales. We plan to synthesize
time-dependent spectra (all four Stokes parameters) of a star with large-scale
surface inhomogeneities, such as active regions, star spots etc. Eventually, we
aim at a comparison to observational data and the calibration or correction of
existing methods to detect and characterize magnetic fields and their global
and local organization on cool main-sequence stars.

\acknowledgments{
The authors acknowledge research funding by the Deutsche Forschungsgemeinschaft (DFG) under the grant SFB 963/1, project A16.
}

\end{document}